\documentclass[letterpaper, 10 pt, conference]{ieeeconf}  % Comment this line out if you need a4paper

\IEEEoverridecommandlockouts                              % This command is only needed if 
\usepackage{amsmath} % assumes amsmath package installed, equation environment
\usepackage{amssymb}  % assumes amsmath package installed
\usepackage{mathrsfs} % for 
\usepackage{amsfonts} % for hollow alphabet
\usepackage{cite}
\usepackage{graphicx}   % dealing with figures
\usepackage{cases}  % for numcases (brace)
\usepackage{xcolor}
\newtheorem{remark}{Remark}

\newtheorem{theorem}{Theorem}
\newtheorem{proposition}{Proposition}
\newtheorem{assumption}{Assumption}

\newcommand{\bea}{\begin{eqnarray}}
\newcommand{\eea}{\end{eqnarray}}

\newcommand{\beas}{\begin{eqnarray*}}
	\newcommand{\eeas}{\end{eqnarray*}}

\def\CR{{\mathcal R}}

\def\CD{{\mathcal D}}

\title{\LARGE \bf
Robust Tracking Control for Nonlinear Systems: Performance optimization via extremum seeking
}

\author{Jiapeng Xu$^{1}$, Ying Tan$^{2}$, and Xiang Chen$^{1}$% <-this % stops a space
\thanks{$^{1}$Jiapeng Xu and Xiang Chen are with the Department of Electrical and Computer Engineering, University of Windsor, Windsor, ON N9B 3P4, Canada.
        {\tt\small jxu@uwindsor.ca, xchen@uwindsor.ca}}%
\thanks{$^{2}$Ying Tan is with the Department of Mechanical Engineering, The
University of Melbourne, Melbourne, VIC 3010, Australia.
        {\tt\small yingt@unimelb.edu.au}}%
}

\begin{document}

\maketitle
\thispagestyle{empty}
\pagestyle{empty}

%%%%%%%%%%%%%%%%%%%%%%%%%%%%%%%%%%%%%%%%%%%%%%%%%%%%%%%%%%%%%%%%%%%%%%%%%%%%%%%%
\begin{abstract}
This paper presents a controller design and optimization framework for nonlinear dynamic systems to track a given reference signal in the presence of disturbances when the task is repeated over a finite-time interval.
This novel framework mainly consists of two steps. The first step is to design a robust linear quadratic tracking controller based on the existing control structure with a Youla-type filter $\tilde Q$.
Secondly,  an extra degree of freedom: a parameterization in terms of $\tilde Q$,
is added to this design framework. This extra design parameter is tuned iteratively from measured tracking cost function with the given disturbances and modeling uncertainties to achieve the best transient performance. 
The proposed method is validated with simulation placed on a Furuta inverted pendulum, showing significant tracking performance improvement.
\end{abstract}

%%%%%%%%%%%%%%%%%%%%%%%%%%%%%%%%%%%%%%%%%%%%%%%%%%%%%%%%%%%%%%%%%%%%%%%%%%%%%%%%
\section{Introduction}
Robust tracking control of nonlinear systems has been extensively studied in the literature using various robust techniques, such as $H_{\infty}$ control \cite{aliyu2011nonlinear,modares2015h} and sliding mode control \cite{elmali1992robust}. These methods are in general the worst-case design, which would ensure the stability under the worst-case disturbances. 
On the other hand, optimal performance such as a linear quadratic form has been the focus of the design in industry applications. However, it is usually hard to analyze the performance for nonlinear dynamics. One of the key reasons to contribute this difficulty comes from the the analysis tool used in stability analysis. Lyapunov direct method \cite{khalil2002nonlinear} has been used to provide sufficient conditions to guarantee the stability of nonlinear dynamics. The optimal control for nonlinear dynamics requires to solve the Hamilton–Jacobi–Bellman (HJB) equation, which is a nonlinear partial differential equation with respect to a given cost. Solving this HJB equation is computational costly.

In contrast,  both robust and optimal control designs have been extensively investigated for linear time-invariant (LTI) dynamic systems \cite{zhou1996robust,zhou2001new,anderson1989optimal,chen2019revisit}. In particular, a robust controller design with a Youla-type filter $\tilde Q$ \cite{chen2019revisit,he2022dual} has been proposed recently, which is motivated by the generalized internal model control (GIMC) proposed in \cite{zhou2001new}. The robust controller with $\tilde Q$ provides automatic robustness recovery in the linear quadratic Gaussian (LQG)/$H_2$ control \cite{chen2019revisit}. Its key idea is to use the $\tilde Q$ filter to balance the optimal performance without consideration of the disturbance and robust performance using the techniques such as $H_\infty$ control. Since the filter $\tilde Q$ is driven by the residual signal indicating the mismatch between the nominal model and the true system,  $\tilde Q$ is only activated when there exists unmodelled dynamics or external disturbances, such that this kind of controller design can lead to a high performance in the presence of disturbances and uncertainties.  This technique is quite different from the traditional mixed $H_2/H_{\infty}$ control, a trade-off design \cite{limebeer1994nash,doyle1994mixed,chen2001multiobjective}.

This work proposes to utilizes the robust controller with $\tilde Q$ in \cite{chen2019revisit} to systematically design the feedback control for a nonlinear dynamic system via its linearization. The proposed framework is used to generate optimal tracking performance for a class of nonlinear dynamic systems to track a given reference trajectory. More specifically, the proposed framework first  presents a robust linear quadratic tracking (LQT) controller design based on the filter $\tilde Q$. Then by introducing an extra gain factor in terms of $\tilde Q$, which can be treated as the balance between the LQT performance and the robustness with respect to disturbances and uncertainties coming from linearizations and other external signals, 
an updating law is generated to tune this gain factor to minimize the tracking cost in the presence of modeling uncertainties and disturbances. The choice of the gain factor does not affect the local stability properties of the closed-loop nonlinear system, while it improves the tracking performance. In this work, the data-driven extremum seeking (ES) approach \cite{ariyur2003real,killingsworth2006pid,tan2010extremum,tan2018model}, which is a model-free optimization method, is adapted to find this optimal gain factor. Alternatively, other model-free optimization techniques such as reinforcement learning can also be considered \cite{modares2015h,hu2022towards}.

The effectiveness of the proposed framework is validated with simulation placed on a Furuta inverted pendulum. It has been shown that this optimal parameter is dependent on the nonlinear dynamics, the type of the reference trajectories, as well as the type of disturbances. The obtained optimal gain can achieve much better transient tracking performance compared with the standard LQT controller and the robust controllers such as $H_\infty$.

The remainder of this paper is organized as follows. Section \ref{sec:problem} formulates the tracking problem for nonlinear systems of interest. Section \ref{sec:robust LQT} presents a design procedure of LQT controller with $\tilde Q$. Section \ref{sec:ES} further presents a controller design for nonlinear systems where performance is further optimized via ES. Section \ref{sec:experiments} provides simulation results on an inverted pendulum. Finally, Section \ref{sec:conclusion} concludes this work.  

\section{Problem Formulation}\label{sec:problem}
Nonlinear systems of the following form are considered:
\begin{align}
    \dot x&=f(x,u,w), \hspace{0.2in}x(0)\in \CR^n\nonumber\\
 y&=g(x,w),\label{nonlinear_plant}
\end{align}
where $x\in \CR^n$ is the state of system, $y\in \CR^p$ is the measurement output, $u\in \CR^m$ is the control input, and  $w\in \CR^{n_w}$ is a disturbance signal representing external disturbances and/or modeling uncertainties. The nonlinear mappings $f:\CD_X\times \CD_U \times \CD_{W}\rightarrow \CR^n$ and
% $h:\CD_X\times \CD_U\rightarrow \CR^{q}$, 
$g:\CR^n\times \CD_W \rightarrow \CR^p$ are continuously differentiable functions. Here $\CD_X$, $\CD_U$, and  $\CD_W$ are compact sets in $\CR^n$, $\CR^m$, and  $\CR^{n_w}$ respectively. Moreover, it is assumed that $f(0,0,0)=0$, indicating the origin is in the set $\Omega=\CD_X\times \CD_U\times \CD_W$.

The linearization of system (\ref{nonlinear_plant}) around $(x=0, u=0, w=0)$ becomes the following LTI system:
\begin{align}
	\dot x &= Ax + B_1w+ B_2u,\label{sys:x}\\
% 	z&=C_1x+D_{12}u,\label{sys:z}\\
	y&=C_2x+D_{21}w,\label{sys:y}
\end{align}
where 
\begin{align*}
    A&=\frac{\partial f}{\partial x}\bigg |_{x,u,w=0},\; B_1=\frac{\partial f}{\partial w}\bigg |_{x,u,w=0},\; \; B_2=\frac{\partial f}{\partial u}\bigg |_{x,u,w=0},\\
    C_2&=\frac{\partial g}{\partial x}\bigg |_{x,w=0},\; \; D_{21}=\frac{\partial g}{\partial w}\bigg |_{x,w=0}.
\end{align*}
The following assumption is standard to stabilize the linearized system (\ref{sys:x}) by output feedback using the output signal in (\ref{sys:y}). 
\begin{assumption}\label{assumption_stabilizable_detectable}
(i) $(A,B_2)$ is stabilizable, and (ii) $(C_2,A)$ is detectable.
\end{assumption}

The control objective of this work is to design an appropriate controller such that the output 
\begin{align}\label{sys:tildey}
    \tilde y=Ex, \; E\in\CR^{p_1\times n},
\end{align}
can track a desired (reference) trajectory $r$ that is known \emph{a priori}. Specifically, we seek to find an optimal control input to minimize the following finite-time quadratic cost function:
\begin{align}\label{sys:cost}
	J(u)=\frac{1}{T}\int_{0}^{T}[(\tilde y-r)'Q(\tilde y-r)+u'Ru] {\rm d}t,
\end{align}
where $Q$ is positive semi-definite and $R$ is positive definite. In this paper, we use ``$'$'' to represent the transpose of a matrix.

We propose to solve the formulated tracking problem for nonlinear systems (\ref{nonlinear_plant})  by two steps. The first step is to design a robust linear quadratic tracking (LQT) controller based on a Youla-type filter $\tilde Q$ proposed in \cite{chen2019revisit} , which will be detailed in the next section. The second step is to introduce
% a parameterization that can balance between the optimal tracking performance and the robustness with respect to disturbances.
an extra gain factor to the filter $\tilde Q$ and use the data-driven optimization algorithm such as extremum seeking (ES) \cite{ariyur2003real,killingsworth2006pid,tan2010extremum,tan2018model} to tune this factor.  

\section{Robust Linear Quadratic Tracking Control with $\tilde Q$}\label{sec:robust LQT}

This section presents a robust LQT controller design based on the filter $\tilde Q$ in \cite{chen2019revisit}, which balances the robustness with respect to disturbances and the LQ optimal tracking control, for the LTI systems taking the same form as the linearized system (\ref{sys:x}) and (\ref{sys:y}).

\subsection{Preliminaries of the controller structure with $\tilde Q$}

\begin{figure}[!htb]
	\centering
	\includegraphics[scale=0.4]{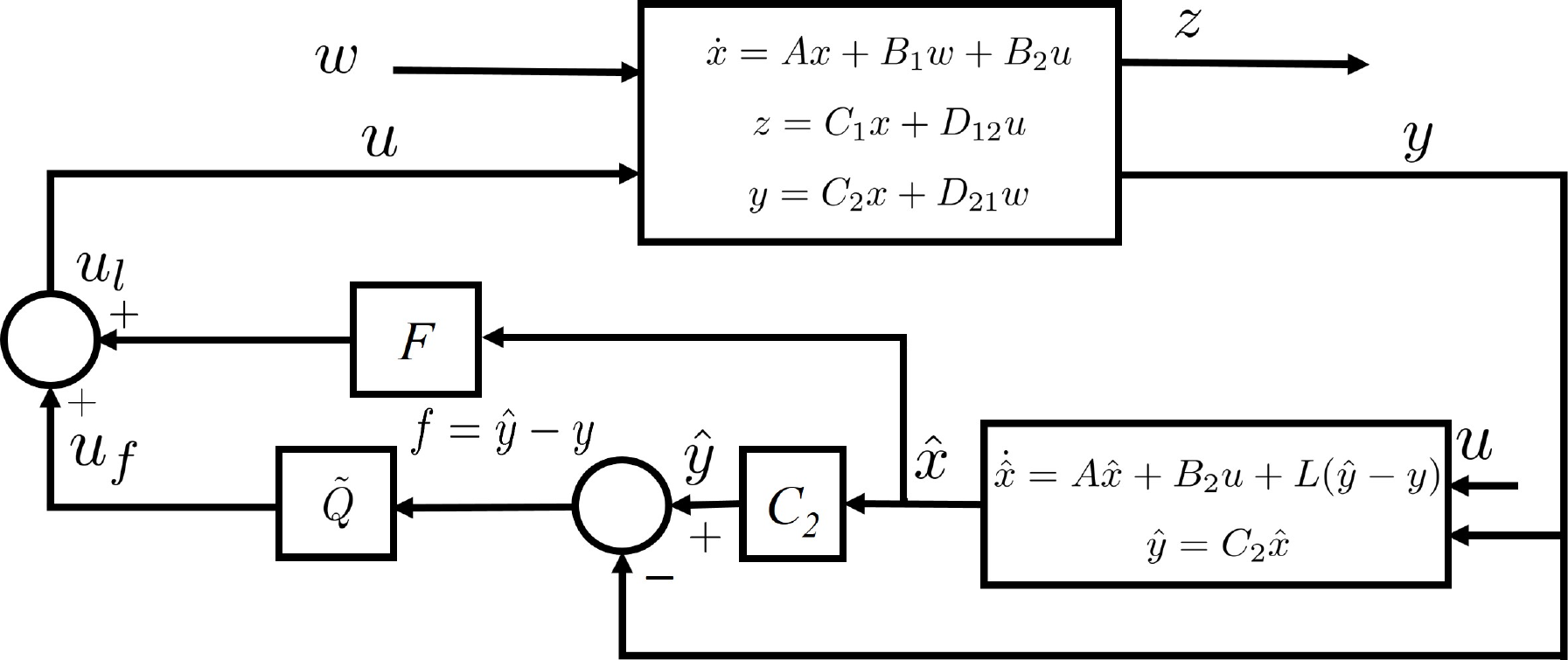}
	\caption{The diagram of the controller with $\tilde Q$.}
	\label{fig:MOCC}
\end{figure}

The structure of the controller with $\tilde Q$ is presented in Fig. \ref{fig:MOCC} (see more discussions in \cite{chen2019revisit}). It contains some performance related variable $z$ defined as
\begin{align}\label{sys:z}
	z=C_1x+D_{12}u,
\end{align}
an observer, and the filter $\tilde Q(\cdot)$ satisfying $\tilde Q(0)=0$.  The input of the filter $\tilde Q$ is the residual signal
$f = \hat y-y$, which is the deviation of estimated and actual sensor outputs. It reflects the mismatch between the nominal model and the true system. As shown in \cite{chen2019revisit}, the output of the filter $\tilde Q(\cdot)$ is designed for robustness recovery, reducing the effect of disturbances. For example, the $H_{\infty}$ design method can be used. 

The control input $u$ consists of two parts : $u_l$ and $u_f$.  The first part $u_l$ is a nominal control, which is to design the stabilizing controller with the help of the observer without consideration of $w(t)$. The matrices $(F,L)$ are the state feedback and observer gains, respectively, such that $A+B_2F$ and $A+LC_2$ are Hurwitz by Assumption \ref{assumption_stabilizable_detectable}. 
The second part $u_f$ is the output of the filter, i.e., $u_f=\tilde Q(f)$.  If $w(t)=0,\forall t\geq 0$, the residual signal $f$ is zero and produces $u_f = \tilde Q(f) = 0$. Hence the role of this filter is to enhance the robustness.
% In the design of $u_f$, the stabilizing controller can be ignored, and only the robustness is the focus. 
One admissible filter $\tilde Q$ is of the following form \cite{chen2019revisit}:  
\begin{align}\label{Qtilde}
	\tilde Q: \; \dot x_{q}=A_qx_q+B_qf,\; \; u_f=F_qx_q.
\end{align}
Matrices $A_q$, $B_q$ and $F_q$ can be obtained by the standard $H_{\infty}$ techniques \cite{doyle1989state,zhou1996robust} for the augmented system with state $(x,e)$, with $e=x-\hat x$, and  $A_q$ is Hurwitz.

Following this controller structure, next will introduce the LQT controller design, followed by the robust controller using $\tilde Q$.

\subsection{Linear Quadratic Tracking (LQT)}
Consider the nominal case of the LTI system (\ref{sys:x}) and (\ref{sys:y}), i.e., $w=0$ and the initial state $x(0)$ is known:
\begin{align}\label{sys:nominal_x}
    \dot x&=Ax+B_2u, \; x(0)~\text{given}.
\end{align}
This reduces to a state feedback control problem.
The following proposition provides the optimal solution to minimizing  (\ref{sys:cost}) by using the Hamilton-Jacobi-Bellman equation \cite[Chapter 2]{anderson1989optimal}; see also \cite[Chapter 4]{anderson1989optimal}.

\begin{proposition}\label{prop:LQT}
Considering the system consisting of (\ref{sys:nominal_x}) and (\ref{sys:tildey}), for a given reference trajectory $r$ and the given performance index (\ref{sys:cost}), the optimal control $u^*$ is given by
	\begin{align}
		-\dot b&=(A+B_2F)'b+E'Qr,\; b(T)=0,\label{b:diff}\\
		u^*&={Fx}+{R^{-1}B_2'b},\label{u_optimal}
	\end{align}
	where $F=-R^{-1}B_2'P$, and $P$ is the solution of the following differential Riccati equation:
    \begin{align}
	-\dot P=PA+A'P-&PB_2R^{-1}B_2'P+E'QE.\label{diff:control}
    \end{align}
\end{proposition}

\begin{figure}[!ht]
	\centering
	\includegraphics[scale=0.5]{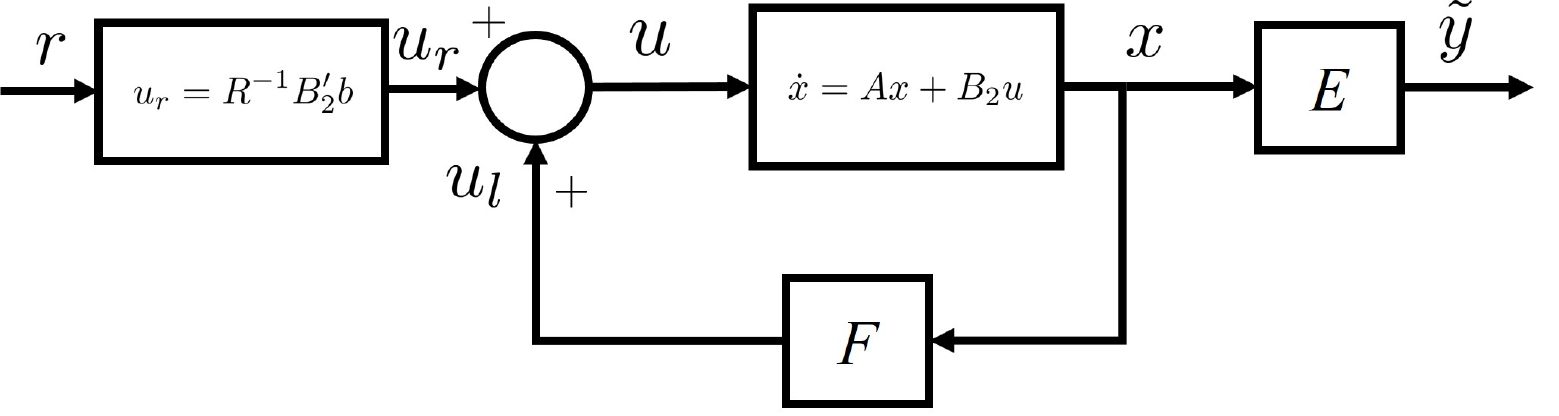}
	\caption{The diagram of LQ optimal tracking control.}
	\label{fig:LQT}
\end{figure}

In practice, to simplify the design of LQT controller, the algebraic Riccati equation instead of differential Riccati equation (\ref{diff:control}) can be considered, such that a time-invariant stabilizing control gain $F$ is obtained. For convenience, the feed-forward control law in (\ref{u_optimal}) is denoted as $u_r:=R^{-1}B_2'b$.
The diagram of control system with LQ optimal tracking is shown in Fig. \ref{fig:LQT}.

\subsection{Design of LQT with $\tilde Q$}

Now we design the LQT controller with $\tilde Q$ to robustify the feedback system. To do this, we redefine the performance variable $z$ in (\ref{sys:z}) as
\begin{align}\label{sys:newz}
	z=C_1(\tilde y-r)+D_{12}u,
\end{align}
such that it coincides with the tracking cost in (\ref{sys:cost}). We can let $C_1'C_1=Q$, $D_{12}'D_{12}=R$ and $C_1'D_{12}=0$ to reduce the weighing matrix parameters.
To design the filter $\tilde Q$, set the reference signal $r$ to be zero such that $u_r=0$ and introduce an observer gain $L$ such that $A+LC_2$ is Hurwitz. Then the design procedure of $\tilde Q$ can be the same as in \cite{chen2019revisit}, i.e., (\ref{Qtilde}). More specifically, the filter $\tilde Q$ is designed according to the following $H_{\infty}$ performance criterion:
\begin{align}\label{hinf}
	\|T_{zw}(s)\|_{\infty}<\gamma,
\end{align}
where $\gamma>0$ is a prescribed value  and $T_{zw}(s)$ is the closed-loop transfer function from $w$ to $z$. See \cite{chen2019revisit} for more details. 
% To solve the $H_{\infty}$ problem, the following assumption is required (see \cite{chen2019revisit} for more details).
% \begin{assumption}\label{assumption_robustness}
% $\left[\begin{array}{ccc}
% 	A-j\omega I & B_2\\
% 		C_1 & D_{12} 
% 	\end{array}\right]$ has full column rank for all $\omega$.
% \end{assumption}

\begin{figure}[!ht]
	\centering
	\includegraphics[scale=0.35]{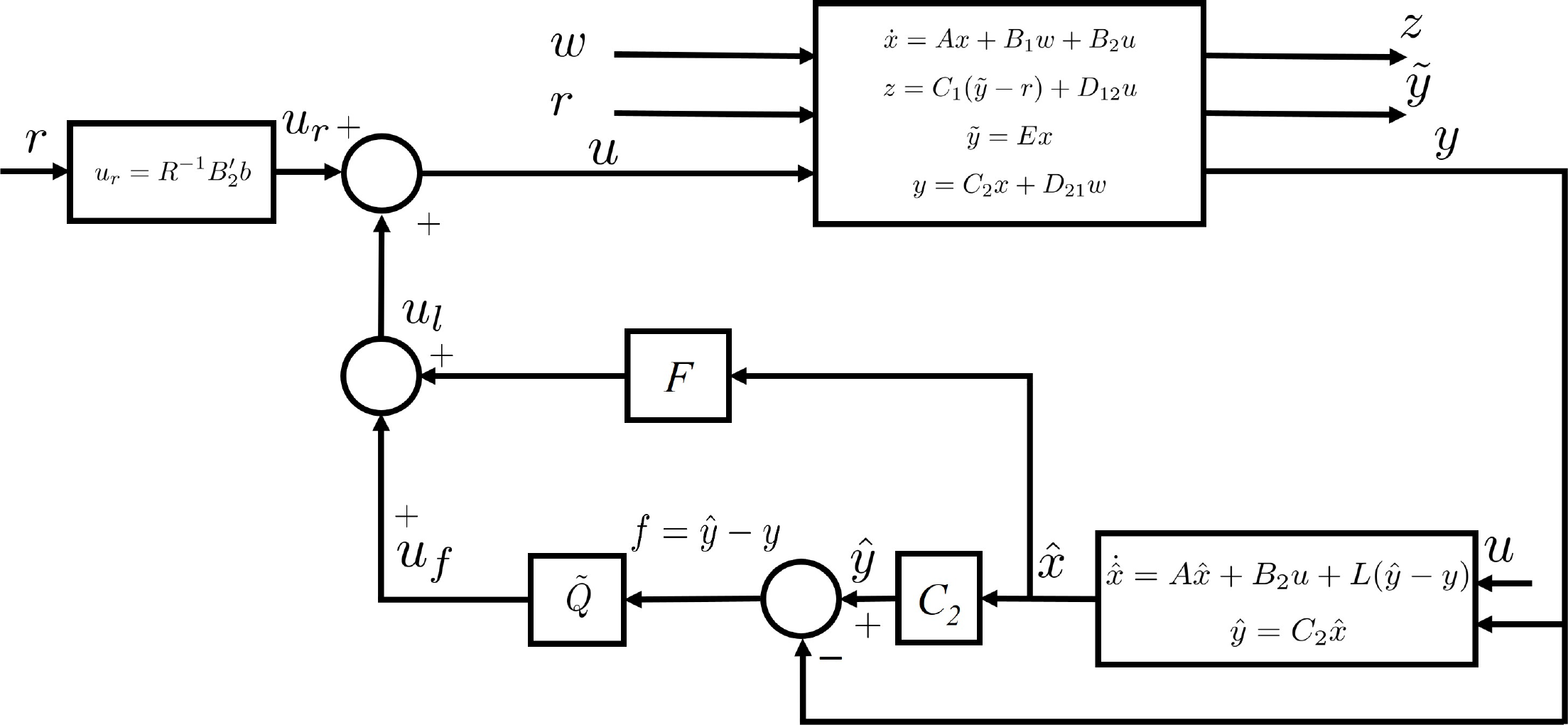}
	\caption{The diagram of robust LQT with parameterized $\tilde Q$.}
	\label{fig:LQT_MOCC}
\end{figure}

By combing the LQT controller, the observer gain $L$ and  the filter  $\tilde Q$ designed using $H_\infty$ technique, the diagram of the robust LQT with $\tilde Q$ is shown in Fig. \ref{fig:LQT_MOCC}.
\begin{remark}
	The transfer functions from $u_r$ to $u$ ,$y$, and $\tilde y$ in the closed-loop system are independent of the filter $\tilde Q$. These transfer function are the same as those in the LQT design. Therefore, the proposed tracking design scheme allows us to select $\tilde Q$ to attenuate disturbances without affecting the tracking capability in the LQT design.
\end{remark}

\section{Performance Optimization for Nonlinear Systems via Extremum Seeking}\label{sec:ES}
We propose to solve the formulated optimal tracking problem  based on the linearized model (\ref{sys:x}) and (\ref{sys:y}). However, the performance of the robust LQT controller with $\tilde Q$ designed in the previous section cannot be guaranteed for nonlinear systems due to the existence of linearization errors. Moreover, this linear controller cannot give an optimal control input  minimizing the performance index (\ref{sys:cost}) for our tracking problem for nonlinear systems (\ref{nonlinear_plant}). In the sequel, we will design an optimal control input by introducing an extra gain factor $\alpha\in \CR$ to the filter $\tilde Q$ and then optimize this factor by the data-driven ES algorithm. 

\subsection{An extra gain factor for $\tilde Q$}
\begin{figure}[!ht]
	\centering
	\includegraphics[scale=0.35]{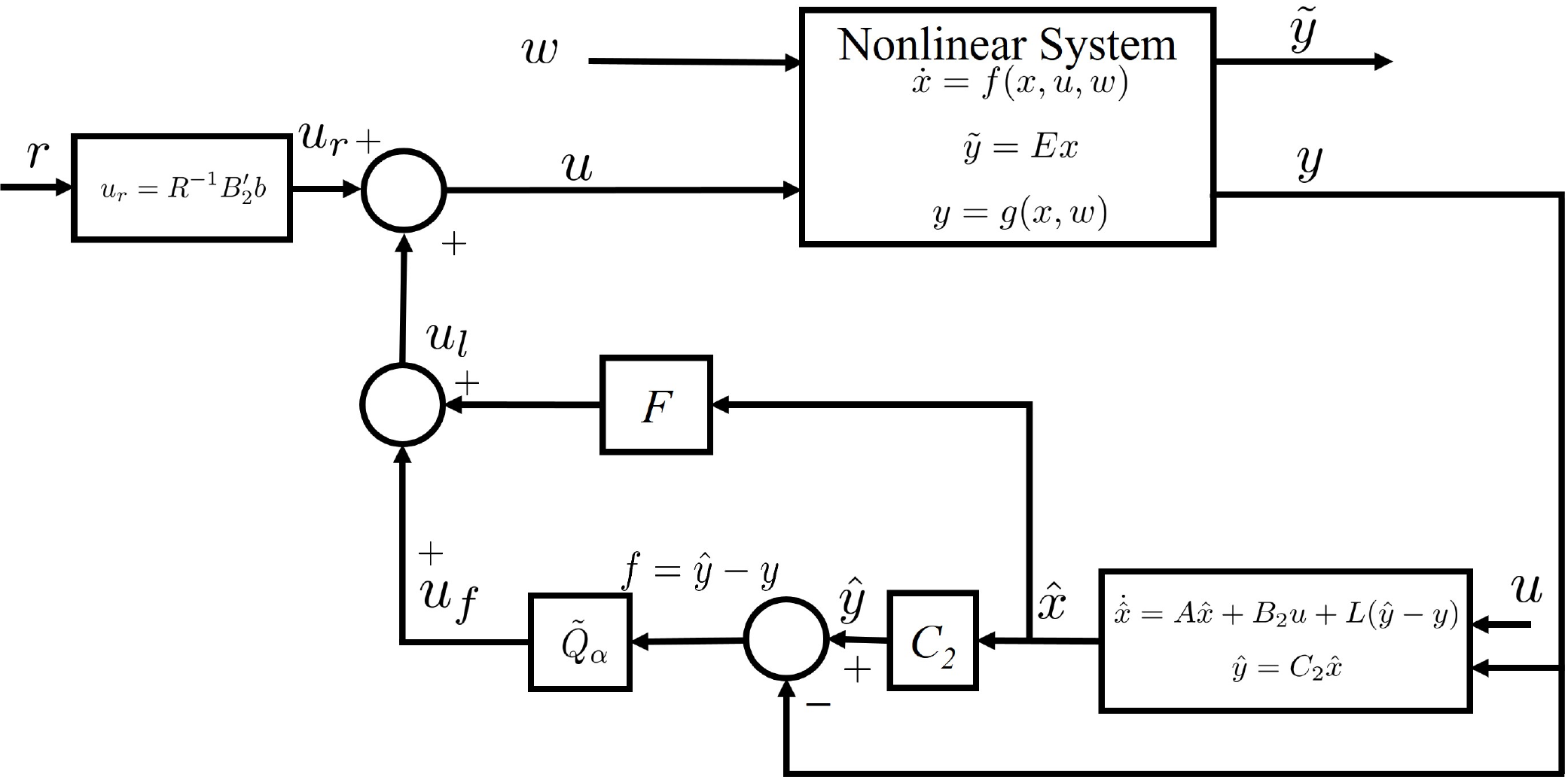}
	\caption{The diagram of robust LQT control with parameterized $\tilde Q_\alpha$ for nonlinear systems.}
	\label{fig:LQT_MOCC_nonlinear}
\end{figure}

To deal with linearization errors and unmodeled disturbances, a gain factor $\alpha\in \CR$ for the output of the filter $\tilde Q$ is introduced as an extra degree of freedom, such that a new $\tilde Q$ denoted by $\tilde Q_{\alpha}$ is obtained:
\begin{align}\label{Qtilde_alpha}
	\tilde Q_{\alpha}: \; \dot x_{q}=A_qx_q+B_qf,\; \; u_f=\alpha F_qx_q.
\end{align}
When $\alpha=0$, it leads to an optimal tracking controller as indicated in Proposition \ref{prop:LQT}  while  $\alpha=1$ leads to the LQT controller with $\tilde Q$ shown in Fig. \ref{fig:LQT_MOCC}. Intuitively, the choice of this $\alpha$ reflects some balance between the LQT and the robustness. This leads to a closed-loop nonlinear system with the LQT controller with $\tilde Q_{\alpha}$ shown in Fig. \ref{fig:LQT_MOCC_nonlinear}. 

% It is highlighted that since the goal of tracking a desired
% output reference is guaranteed by the feedforward controller $u_r$ in Fig. \ref{fig:LQT}, it is convenient to linearize the nonlinear system only at the origin ($x=0,u=0,w=0$), leading to a LTI system as in (\ref{sys:x}) and (\ref{sys:y}).  Hence 
% %As the linearization of system (\ref{nonlinear_plant}) around $(x=0,u=0,w=0)$ becomes a LTI system shown in (\ref{sys:x}) and (\ref{sys:y}), 
% the proposed the robust tracking controller, i.e., LQT with $\tilde Q$ developed in the previous section is applicable for the the nonlinear system when the controller is designed via the linearized model consisting of (\ref{sys:x}) and (\ref{sys:y}).

Now the tracking performance index $J(u)$ in (\ref{sys:cost}) can be written as a function of $\alpha$, i.e., 
\begin{align}\label{es:cost}
    J(\alpha)=\frac{1}{T}\int_{0}^{T}[(\tilde y-r)'Q(\tilde y-r)+u'Ru] {\rm d}t,
\end{align}
and we seek to find an optimal $\alpha$ parameter to minimize the above cost function. Before solving the $\alpha$ optimization problem, we present the following result about the local stability properties of such a closed-loop nonlinear system.
% Basically, we can use its linearization to design appropriate controllers while the linearization error can be treated as a part of the disturbance.  

\begin{theorem}\label{thm:alpha}
	Let the origin ($x=0,u=0,w=0$) be an equilibrium point for the nonlinear system (\ref{nonlinear_plant}) and let system (\ref{sys:x}) and (\ref{sys:y}) be the linearization of (\ref{nonlinear_plant}) about the origin. Then for any $\alpha\in \CR$, the origin of the closed-loop nonlinear system in Fig. \ref{fig:LQT_MOCC_nonlinear} is locally exponentially stable.
\end{theorem}

\proof 
Let $x_c:=\left[\begin{array}{cc}
	x_{ q}' & \hat x'
\end{array}\right]'$. Then it follows from the design of the LQT controller with $\tilde Q$ shown in Fig. \ref{fig:LQT_MOCC} and the expression of $\tilde Q_{\alpha}$ in (\ref{Qtilde_alpha}) that we have the following linear tracking controller:
\begin{align}\label{xc}
    \dot x_c &=A_cx_c+B_cy+B_ru_r,\; u=F_cx_c+u_r,
\end{align}
where 
\begin{align*}
    A_c&=\left[\begin{array}{cc}
		A_{ q} &B_{ q}C_2 \\
		\alpha B_2F_q & A+B_2F+LC_2
	\end{array}\right],\; B_c=\left[\begin{array}{c}
		-B_{ q} \\
		-L
	\end{array}\right], \\
	B_r&=\left[\begin{array}{c}
	0 \\ B_2
	\end{array}\right],\; F_c=\left[\begin{array}{cc}
	\alpha F_q & F
	\end{array}\right].
\end{align*}
Hence, by a linear transformation, it can be shown that the closed-loop matrix for the linearized system (\ref{sys:x}) and (\ref{sys:y}) has the following triangular form
\begin{align}\label{TAT}
	\bar A=\bar A(\alpha)=\left[\begin{array}{ccc}
		A+B_2F & \alpha B_2F_q &B_2F\\
		0 & A_{ q} &B_{ q}C_2 \\
		0 & 0 & A+LC_2
	\end{array}\right].
\end{align}
Since $A+B_2F$, $A+LC_2$ and $A_q$ are all Hurwitz, $\bar A$ is Hurwitz. It can be verified that the origin $(x = 0, x_c = 0, w=0)$ is an equilibrium point of the
closed-loop nonlinear system. Thus, we can conclude the local exponential stability of the closed-loop nonlinear system in Fig. \ref{fig:LQT_MOCC_nonlinear} at the origin for any $\alpha\in \CR$\cite[Section 12.2]{khalil2002nonlinear}. \hfill\rule{2mm}{2mm}

Theorem \ref{thm:alpha} shows that the choice of $\alpha$ does not affect the local stability properties of the closed-loop nonlinear system presented in Fig. \ref{fig:LQT_MOCC_nonlinear} while it affects the performance, in particular, the transient performance.  This work tries to find an optimal $\alpha$ to tracking control for nonlinear systems. In the sequel, the data-driven extremum seeking (ES) \cite{ariyur2003real,killingsworth2006pid,tan2010extremum,tan2018model} will be used tune the parameter $\alpha$ for a given disturbance $w$.

\subsection{Performance optimization via extremum seeking}
It is highlighted that in the linearized model consisting of (\ref{sys:x}) and (\ref{sys:y}), the disturbances coming from two parts: one is unmodelled uncertainties coming from the linearization residue and unmodelled dynamics. Such uncertainties are related to the size of compact sets $\CD_X$, $\CD_U$, and $\CD_W$. The other is other types of deterministic and random noises.  In this paper, our focus is the unmodelled uncertainties and deterministic noises that are repeatable when the nonlinear dynamics run over a fixed time interval $[0,T]$. Thus, the disturbance can be re-written as $w=w(t,x,u)$. 

\begin{assumption}\label{assumption_ESC}
It is assumed that for a deterministic and repeatable disturbance $w(t,x,u), t\in [0,T]$, for any $x\in \CD_X$, and $u\in \CD_U$, there exists a unique optimal $\alpha^*\in \CR$ such that the cost function defined (\ref{es:cost}) can reach a minimum. 
\end{assumption}

\begin{figure}[!ht]
	\centering
	\includegraphics[scale=0.35]{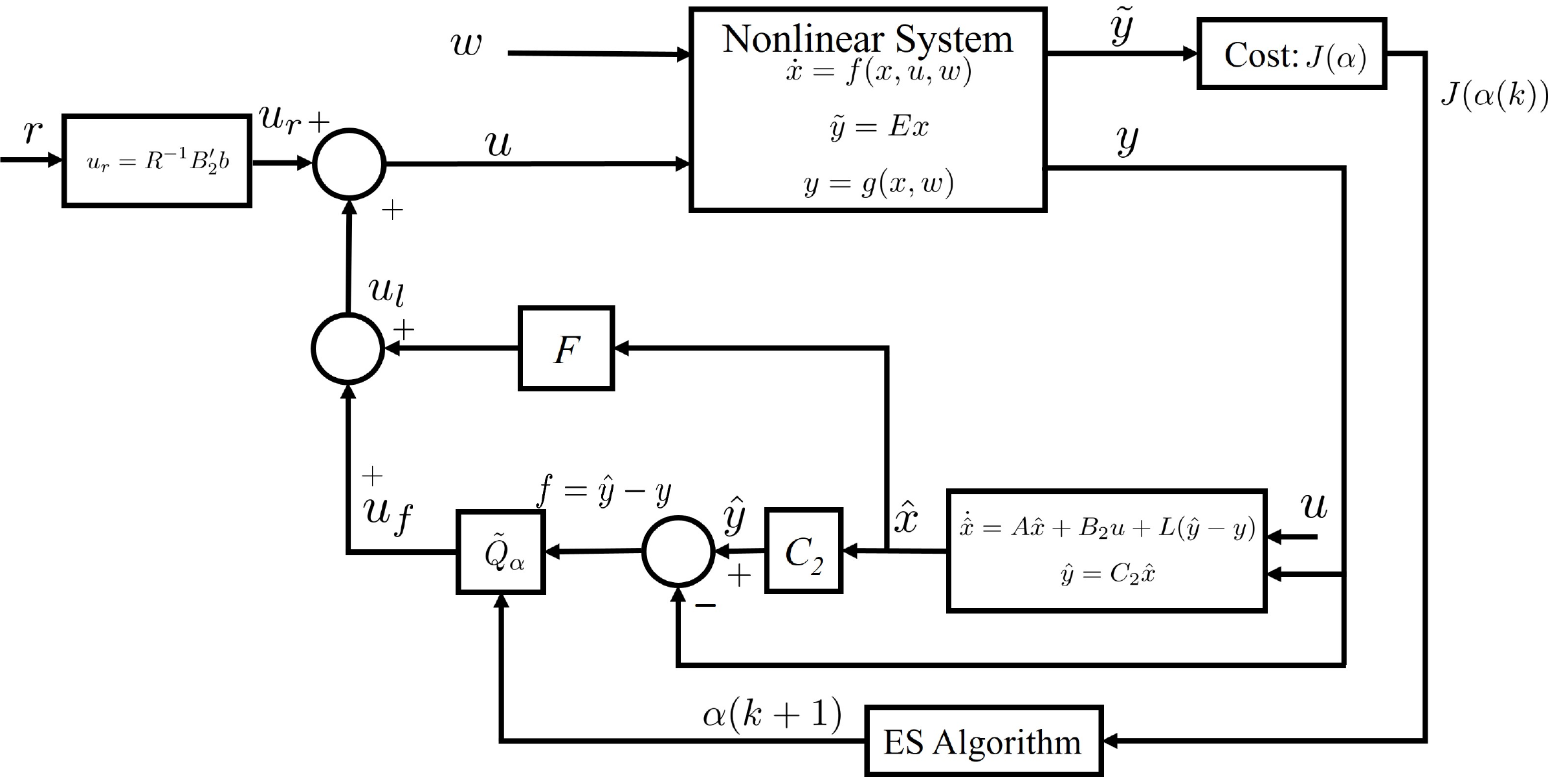}
	\caption{The overall ES $\alpha$ tuning scheme. The ES algorithm updates the parameter $\alpha(k)$ in the filter $\tilde Q_{\alpha}$ to minimize the cost $J(\alpha)$.}
	\label{fig:ES}
\end{figure}

Under this assumption,  the ES algorithm presented in \cite{ariyur2003real}  is used to  tune the parameter $\alpha$ for a given disturbance $w(t,x,u)$ by repeatedly running the closed-loop nonlinear system consisting of (\ref{nonlinear_plant}) with the control law (\ref{xc}) over the finite time $[0,T]$.
ES is a model-free optimization method which uses only input–output data  to see an optimal input with respect to a given cost \cite{ariyur2003real}. The ES algorithm adopted here works for the iteration domain, which is the same as the one used in \cite{killingsworth2006pid}: 
\begin{align}
	\zeta(k+1)&=-h\zeta(k)+J(\alpha(k)),\notag\\
	\hat\alpha(k+1)&=\hat\alpha(k)-\delta\beta\cos(\omega k)[J(\alpha(k))-(1+h)\zeta(k)],\notag\\
\alpha(k+1)&=\hat\alpha(k+1)+\beta\cos(\omega(k+1)),\label{es}
\end{align}
where $k$ is the iteration number, $0<h<1$, $\zeta(k)$ is a scalar, $\delta$ is the step size, and $\beta$ is the perturbation amplitude. Stability and convergence are mainly influenced by the values of $\delta$ and $\beta$. The modulation frequency $\omega$ is chosen such that $\omega=a\pi$, where $a$ satisfies $0< a < 1$. The overall ES $\alpha$ tuning scheme is summarized in Fig. \ref{fig:ES}. The local convergence analysis of such ES algorithm when Assumption \ref{assumption_ESC} holds locally can be found in \cite{choi2002extremum}. Using the similar analysis techniques as in \cite{tan2010extremum} for the continuous-time systems, non-local convergence can be achieved when Assumption \ref{assumption_ESC} holds.

The design procedure of the proposed robust optimal tracking control scheme can be summarized as the following three steps.
\begin{itemize}
    \item Step 1: Obtain the linearized model around the origin $(x=0,u=0,w=0)$ from nonlinear system (\ref{nonlinear_plant});
    \item Step 2: Design the LQT controller with $\tilde Q_{\alpha}$ as in Fig. \ref{fig:LQT_MOCC} in which $\tilde Q$ in (\ref{Qtilde}) is replaced by $\tilde Q_{\alpha}$ in (\ref{Qtilde_alpha});
    \item Step 3: Tune $\alpha$ via the ES algorithm (\ref{es}) for a given disturbance $w(t,x,u)$, shown in  Fig. \ref{fig:ES}. 
\end{itemize}

\begin{remark}
It is noted that the focus of this work is to ensure that the transient behaviour of the closed-loop system presented in Fig. \ref{fig:ES} over the finite time interval $[0,T]$ can be improved over iteration when the disturbance $w$ is repeatable over iteration. If Assumption \ref{assumption_ESC} holds, the ES algorithm in (\ref{es}) can ensure that the cost function (\ref{es:cost}) decreases over iterations for appropriately tuned parameters $(a,\delta, \beta, h)$ in (\ref{es}). The decrease of the cost function has been verified in simulations.  
\end{remark}

\section{Simulation Results}\label{sec:experiments}
In order to verify the effectiveness of the proposed robust tracking control in Fig. \ref{fig:ES}, a simulation example on a Furula inverted pendulum (the Quanser QUBE Servo 2 rotary pendulum \cite{apkarianqube}) is presented. 

\subsection{Linearization and design parameters}

The  Furuta inverted pendulum  is an under-actuated (unstable) nonlinear system with the state variable $x =\left[\begin{array}{cccc}
	\theta_1 &   \theta_2 &   \dot\theta_1   &    \dot\theta_2
\end{array}\right]'$, consisting of the base arm angle
$\theta_1$ and the vertical
arm (pendulum) angle $\theta_2$, as well as the corresponding angular velocities. When the pendulum is in the upright position, $x =\left[\begin{array}{cccc}
	0 &   0 &  0   &   0
\end{array}\right]'$.  
% The equations of motion (EOM) for the pendulum system are developed using the Euler-Lagrange method. The resultant nonlinear EOM
The nonlinear equations of motion for the pendulum system in disturbance-free case are \cite{apkarianqube}:
\begin{align}\label{nonlinear_pendulum}
	M\ddot\theta_1=&m_plr\ddot\theta_2\cos\theta_2-J_p\dot\theta_1\dot\theta_2\sin2\theta_2-m_plr\dot\theta_2^2\sin\theta_2\notag\\
	&-b_r\dot\theta_1+\tau,\notag\\
	J_p\ddot\theta_2=&m_plr\ddot\theta_1\cos\theta_2+0.5J_p\dot\theta_1^2\sin2\theta_2+m_pgl\sin\theta_2\notag\\
	&-b_p\dot\theta_2,
\end{align}
where $M=J_r+J_p\sin^2\theta_2$ and $\tau=k_m/R_m(u-k_m\dot\theta_1)$ is the applied torque at the base of the rotary arm with $u$ the control input/voltage. 

Following the 3-step design procedure in the previous section, we need to linearize the nonlinear system first. The origin ($x=0,u=0$) is an equilibrium point for pendulum system (\ref{nonlinear_pendulum}), leading to the following  linearization model with assumed disturbance/noise $w$  
\begin{align*}
	A&=\left[\begin{array}{cccc}
		0    &     0 &   1   &      0 \\
		0    &     0   &      0 &   1\\
		 0 & \frac{m_p^2l^2rg}{J_t} & -\frac{J_pb_r}{J_t}-\frac{k_m^2J_p}{R_mJ_t} &  -\frac{m_plrb_p}{J_t}\\ 
		0 & \frac{J_rm_pgl}{J_t} &   -\frac{m_plrb_r}{J_t}-\frac{k_m^2m_plr}{R_mJ_t} &  -\frac{J_rb_p}{J_t}
	\end{array}\right],\\
	B_2&=\left[\begin{array}{c}
		0\\
		0\\
		\frac{k_mJ_p}{R_mJ_t}\\
		\frac{k_mm_plr}{R_mJ_t}
	\end{array}\right],\;
	B_1={\rm diag}\{0.012, 0.012, 1, 1\},\\
	C_2&=\left[\begin{array}{cccc}
		1 & 0 & 0 & 0\\
		0 & 1 & 0 & 0
	\end{array}\right],\; D_{21}=\left[\begin{array}{cccc}
		0 & 0 & 0.0001 & 0\\
		0 & 0 & 0 & 0.0001
	\end{array}\right].
\end{align*}
where $J_t=J_pJ_r-m_p^2l^2r^2$. The values of system parameters of the rotary pendulum are presented in Table \ref{table-para}. Since $w$ is considered as linearization errors and deterministic unmodelled disturbances, the disturbance matrices $B_1$ and $D_{21}$ above are chosen by trial and error. Let 
	\begin{align}
	    \tilde y&=\left[\begin{array}{cc}
		1    &     0 
	\end{array}\right]y,
	\end{align}
	such that $E=\left[\begin{array}{cccc}
	1 & 0 & 0 & 0
\end{array}\right]$, meaning that the base arm angle $\theta_1$ needs to track a reference signal and the vertical pendulum needs to be balanced in the upright position. 

\begin{table}[!htbp]
        \scriptsize
		\centering 
		\caption{\centering System parameters of the rotary pendulum.}\label{table-para}
		\begin{tabular}{cccc}
			\hline
			\textbf{Parameter} & \textbf{Description} & \textbf{Value} & \textbf{Unit}\\
			\hline
			$R_m$ & Terminal resistance & 8.4 & $\Omega $\\
			$k_m$ & Motor back-emf constant & 0.042 & V$\cdot$s/rad\\
			$r$ & Length of the base arm & 0.085 & m\\
			$2l$ & Length of the pendulum & 0.129 & m\\
			$m_p$ & Mass of the pendulum  & 0.024 & Kg\\
			$g$ & Gravity constant & 9.81 & N/Kg\\
			$b_r$ & Damping on the rotary arm & 0.0005 & N$\cdot{\rm m}\cdot$ s/rad\\
			$b_p$ & Damping on the pendulum & 0.0001 & N$\cdot {\rm m}\cdot$ s/rad\\
			$J_r$ & Inertia of the rotary arm & $2.3060\times 10^{-4}$ & N$\cdot {\rm m}\cdot {\rm s}^2$/rad\\
			$J_p$ & Inertia of the pendulum & $1.3313\times 10^{-4}$ & N$\cdot {\rm m}\cdot {\rm s}^2$/rad\\
			\hline
		\end{tabular}
\end{table}	

The weighting matrices for the performance variable $z$ in (\ref{sys:z}) and for the tracking performance (\ref{es:cost}) are selected as $C_1=\left[\begin{array}{cc}
		15 & 0\\
	\end{array}\right]', D_{12}=\left[\begin{array}{cccc}
		0 & \sqrt{2}
	\end{array}\right]', R=C_1'C_1$ and $Q=D_{12}'D_{12}$.  
Then according to Step 2, we need to design the LQT controller with $\tilde Q_{\alpha}$. The LQT controller in Proposition \ref{prop:LQT}  is first designed for the 
linearized pendulum system, in which the closed-loop poles are located at ($-16.55\pm j12.80, -21.20\pm j1.76$). The observer gain $L$ is chosen such that the eigenvalues of $A+LC_2$ are ($-59.40\pm j80.54, -61.04\pm j76.24$). The filter $\tilde Q$ is then designed to satisfy the $H_{\infty}$ performance (\ref{hinf}) \cite{chen2019revisit}, where $\gamma=0.21$.
% which is the lowest value the $H_{\infty}$ controller can achieve. 
Following Step 3, we shall tune $\alpha$ via the ES algorithm (\ref{es}) in simulation under different disturbance signals, shown in the next subsection.  

\subsection{Simulation results}
We consider that the disturbance signal $w$ is injected into the control voltage $u$ and the measurement output $y$ is perturbed by white noise. In particular, two different disturbance signals are considered: a square wave with the amplitude of $2$ and frequency of 0.5Hz, denoted by $w_1$, and a kind of vanishing signal $w_2(t)=5e^{-0.1t}$. Two different references are considered: Case A tracks a square wave with amplitude of 20 degrees and frequency of 0.05Hz while Case B tracks a sinusoidal signal $r=\frac{\pi}{3}\sin\pi t$. In the simulation, the initial value of ES is selected as $\alpha(0)=1$, indicating the local $H_\infty$ performance at the first iteration.

For all simulations, the quadratic cost function $J(\alpha)$ in (\ref{es:cost}) is used with $T=20s$ and the parameters $a$, $h$ and $\beta$ in the ES algorithm (\ref{es}) are set to 
$$ a=0.8, h=0.1, \beta=0.015.$$
The choice of the step size $\delta$ in the ES algorithm depends on a specific disturbance signal and reference signal. The optimal $\alpha$ tuned via ES is denoted as $\alpha^*$.

\begin{figure}[!ht]
	\centering
	\includegraphics[scale=0.5]{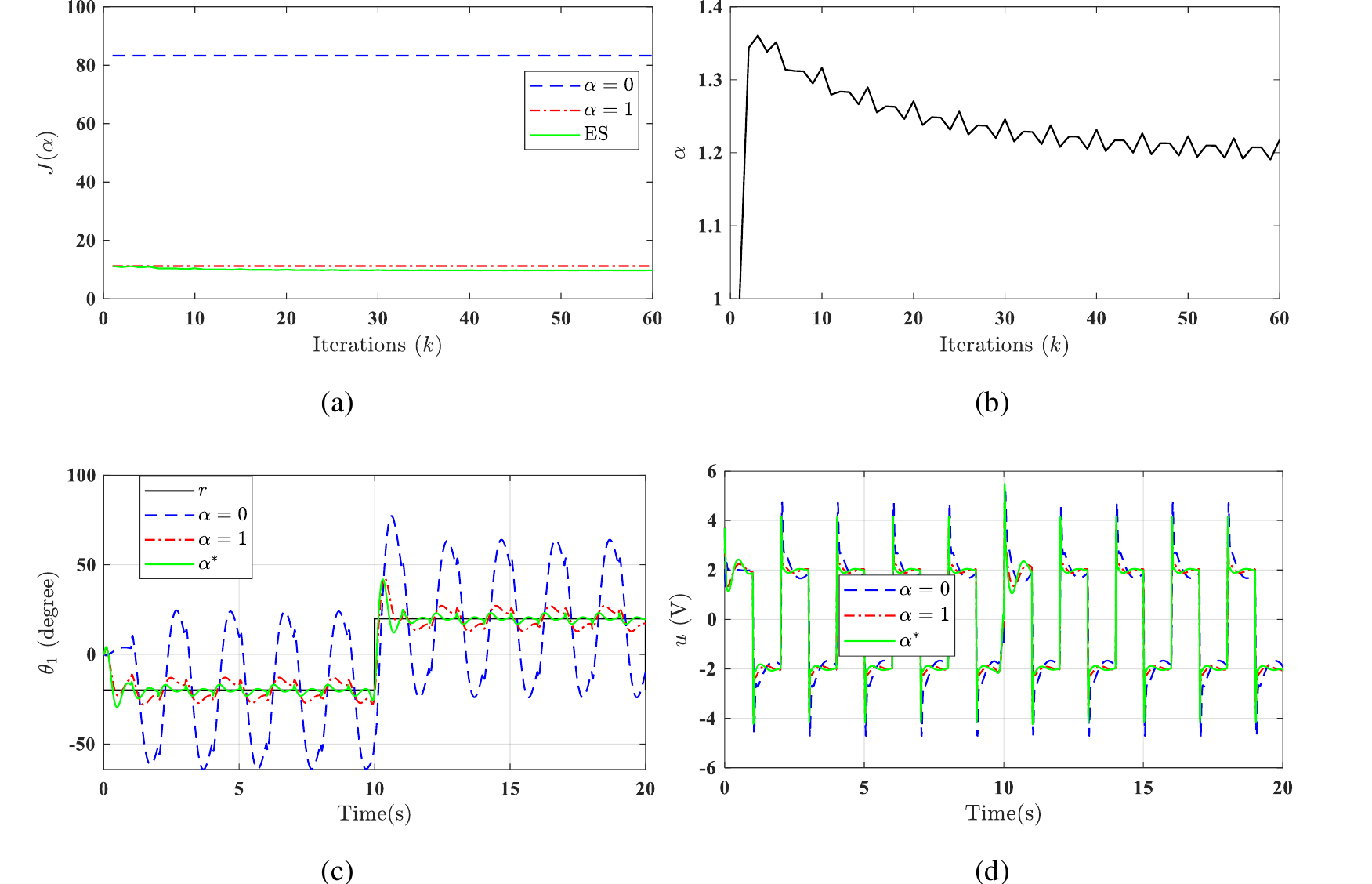}
	\caption{Case A: ES $\alpha$ tuning for $w_1$ illustrated by (a) the evolution of the cost function and (b) the parameter $\alpha$ during ES tuning of the closed-loop system with $w_1$. The lower plots present (c) the output signal $\theta_1$ and (d) the control input signal $u$.}
	\label{fig:ES_w2_caseA}
\end{figure}

\begin{figure}[!ht]
	\centering
	\includegraphics[scale=0.5]{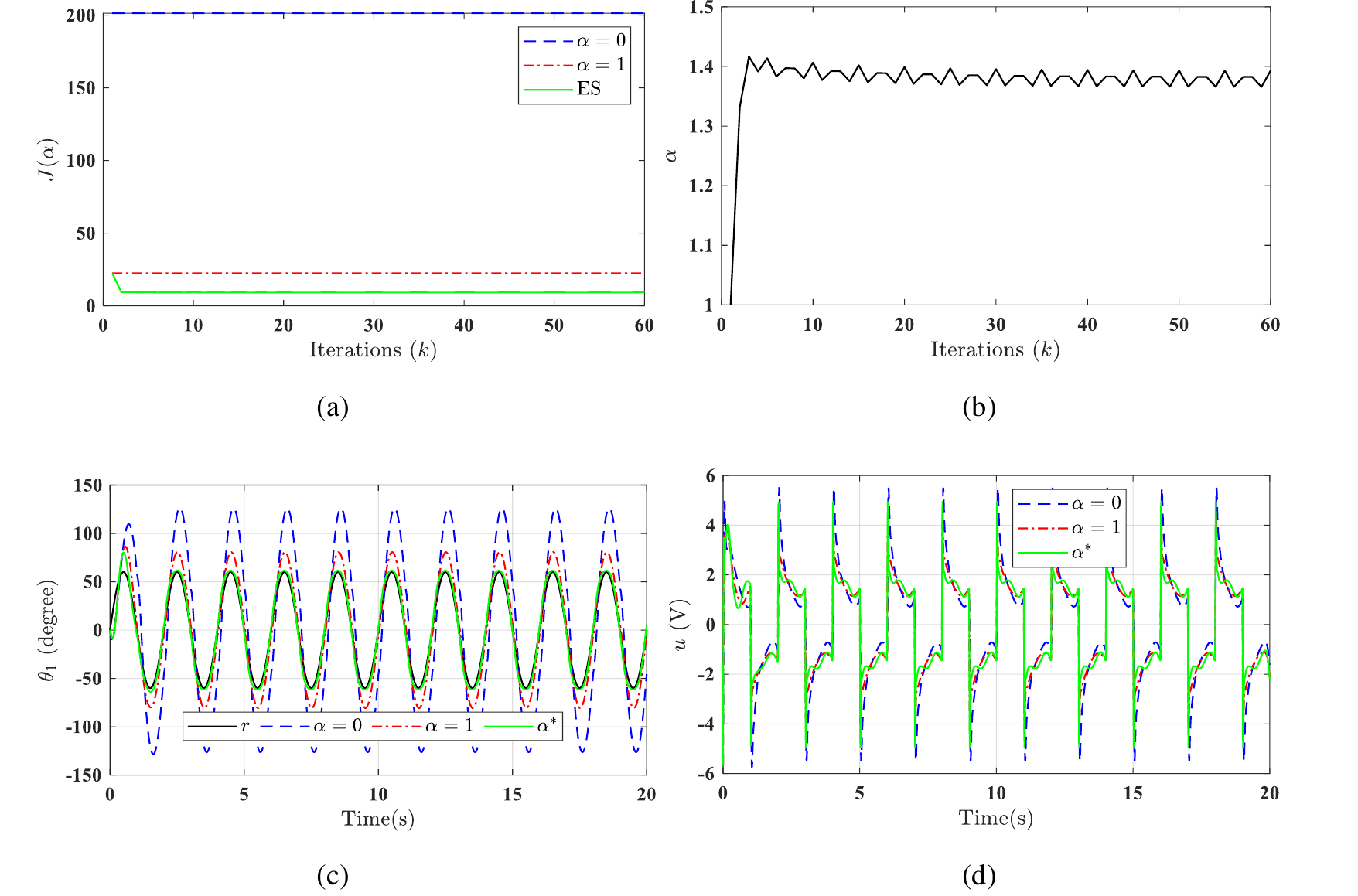}
	\caption{Case B: ES $\alpha$ tuning for $w_1$ illustrated by (a) the evolution of the cost function and (b) the parameter $\alpha$ during ES tuning of the closed-loop system with $w_1$. The lower plots present (c) the output signal $\theta_1$ and (d) the control input signal $u$.}
	\label{fig:ES_w2_caseB}
\end{figure}

Now we tune $\alpha$ for disturbances $w_1$ and $w_2$.  Table \ref{table-w2} summarizes the $\alpha^*$ values in which minimum costs are reached for Case A and Case B in the presence of $w_1$ and $w_2$.  Figs. \ref{fig:ES_w2_caseA} and \ref{fig:ES_w2_caseB} show that ES minimizes the cost function (\ref{es:cost}) with convergence to the parameter $\alpha$ that produces a minimum for Case A and Case B in the presence of $w_1$ respectively. It can be seen that the controller with $\alpha^*$ obtained from ES tuning yields a much better closed-loop performance in terms of transient behaviours compared with the cases when $\alpha=0$ (LQT performance) and $\alpha=1$ ($H_\infty$ performance). Figs. \ref{fig:ES_w2_caseA} and \ref{fig:ES_w2_caseB} show the effectiveness of the proposed ES algorithm when dealing with deterministic uncertainties and linearization errors. Similar results are observed when dealing with $w_2$.

\begin{table}[!htbp]
        \scriptsize
		\centering 
		\caption{\centering $\alpha^*$ for disturbances $w_1$ and $w_2$.}\label{table-w2}
		\begin{tabular}{cccc}
			\hline
			 ~& $w_1$ & $w_2$\\
			 \hline
			Case A & $\alpha^*=1.20$ & $\alpha^*=1.21$ \\
			Case B & $\alpha^*=1.38$ & $\alpha^*=1.47$ \\
			\hline
		\end{tabular}
\end{table}	

\section{Conclusion}\label{sec:conclusion}
In this paper, we develop a novel tracking control scheme for a classe of nonlinear systems in the presence of disturbances based on the robust controller with a Youla-type filter $\tilde Q$ and the data-driven ES technique. A key point is that a gain factor $\alpha$ is introduced to the filter $\tilde Q$, which allows an extra degree of freedom to be optimized for tracking performance. It has been shown from the simulation results that the proposed tracking controller with a tuned $\alpha$ via ES can achieve an optimal tracking performance for nonlinear systems with disturbances. 
% ES used in this work is in an off-line/learning way, which is easy to implement in practice. It would be interesting to explore more appropriate disturbances as ES input data, such that the tuned parameter $\alpha$ can handle various disturbances better.

\addtolength{\textheight}{-12cm}   % This command serves to balance the column lengths

%%%%%%%%%%%%%%%%%%%%%%%%%%%%%%%%%%%%%%%%%%%%%%%%%%%%%%%%%%%%%%%%%%%%%%%%%%%%%%%%

%%%%%%%%%%%%%%%%%%%%%%%%%%%%%%%%%%%%%%%%%%%%%%%%%%%%%%%%%%%%%%%%%%%%%%%%%%%%%%%%

%%%%%%%%%%%%%%%%%%%%%%%%%%%%%%%%%%%%%%%%%%%%%%%%%%%%%%%%%%%%%%%%%%%%%%%%%%%%%%%%
%\section*{APPENDIX}
%\section*{ACKNOWLEDGMENT}
%
%
%
%
%%%%%%%%%%%%%%%%%%%%%%%%%%%%%%%%%%%%%%%%%%%%%%%%%%%%%%%%%%%%%%%%%%%%%%%%%%%%%%%%%
%
%References are important to the reader; therefore, each citation must be complete and correct. If at all possible, references should be commonly available publications.

\bibliographystyle{IEEEtran}
\bibliography{mybibfile}

\end{document}